\documentclass[3p,times]{elsarticle}

\long\def\comment#1{ }
\newcommand{\eqn}[1]{Eq.~(\ref{#1})}
\newcommand{\beq}{\begin{equation}}
\newcommand{\eeq}{\end{equation}}

\newcommand{\dif}{{\rm d}}
\newcommand{\rmd}{{\rm d}}
\newcommand{\rme}{{\rm e}}
\newcommand{\rmi}{{\rm i}}

\newcommand{\rmtr}{{\rm tr}}
\newcommand{\rmTr}{{\rm Tr}}
\newcommand{\del}{\partial}

\newcommand{\mcal}{\mathcal}

\newcommand{\bk}{\bm{k}}

\newcommand{\bp}{\bm{p}}
\newcommand{\bx}{\bm{x}}
\newcommand{\by}{\bm{y}}
\newcommand{\bu}{\bm{u}}

\newcommand{\bz}{\bm{z}}
\newcommand{\bw}{\bm{w}}

\newcommand{\bbx}{\bm{\bar{x}}}
\newcommand{\bby}{\bm{\bar{y}}}

\newcommand{\pd}{{\phantom{\dagger}}}

\usepackage{ecrc}

\volume{00}

\firstpage{1}

\journalname{Nuclear Physics A}

\runauth{E.~Iancu and D.N.~Triantafyllopoulos}

\jid{nupha}

\usepackage{bm,amsmath,amssymb}

\biboptions{comma,square,sort&compress}

\usepackage[figuresright]{rotating}

\begin{document}

\begin{frontmatter}




\title{JIMWLK evolution for multi-particle production with rapidity correlations}


\author[ei]{E.~Iancu}
\ead{edmond.iancu@cea.fr}

\author[dnt]{D.N.~Triantafyllopoulos\corref{cor1}}
\ead{trianta@ectstar.eu}

\cortext[cor1]{Corresponding author}

\address[ei]{Institut de Physique Th\'{e}orique de Saclay, F-91191 Gif-sur-Yvette, France}
\address[dnt]{European Centre for Theoretical Studies in Nuclear Physics and Related Areas (ECT*) and Fondazione Bruno Kessler, \\ Strada delle Tabarelle 286, I-38123 Villazzano (TN), Italy}

\begin{abstract}
We study multi--particle production with rapidity correlations in proton-nucleus collisions at high energy in the Color Glass Condensate framework. The high-energy evolution responsible for such correlations is governed by a generalization of the JIMWLK equation describing the simultaneous evolution of the strong nuclear color fields in the direct amplitude and the complex conjugate amplitude. This functional equation can be used to derive ordinary evolution equations for the cross-sections for particle production, but the ensuing equations appear to be too complicated to be useful in practice, including in the limit of a large number of colors $N_c$. We propose an alternative formulation 
based on a Langevin process, which is valid for generic $N_c$ and is better suited for numerical implementations. For illustration, we present the stochastic equations which govern two gluon production with arbitrary rapidity separation.

\end{abstract}


\end{frontmatter}

\section{Introduction}
\label{sec:intro}

The experimental observation of multi--particle correlations in particle production in proton--proton (p+p), proton--nucleus (p+A), and nucleus--nucleus (A+A) collisions
at RHIC and the LHC provides essential information about the non-linear phenomena associated
with high parton densities. The long-range correlations in pseudo-rapidity $\Delta \eta$ are
particularly interesting in that sense. On one hand, causality arguments suggest that such
correlations must have been built at early times in the collision, so they are likely
to encode information about the hadronic wave functions prior to the collision.
On the other hand, they may be affected by final--state interactions which occur at later stages, and thus be sensitive to collective phenomena, like flow, in the partonic fireball.
For instance, the `ridge' effect seen in A+A collisions at both
RHIC and the LHC is generally interpreted as a combination of initial--state correlations in
rapidity and final--state collective flow leading to azimuthal collimation. Yet, this interpretation
has recently been shaken by the discovery of similar phenomena in p+A and even p+p collisions, in special events with high multiplicity.  Strong final--state effects are
a priori not expected in such collisions, because of the small size
and reduced lifetime of the respective `fireballs'. This invite us to reexamine the 
formation of (rapidity and angular) correlations in the initial state, notably via the high-energy 
evolution. A major difficulty is the lack of factorization for the calculation of
multi--particle production at different rapidities while taking into account multiple scattering. In the context of p+A collisions, we have recently provided an explicit solution to this problem \cite{Iancu:2013uva}, to be reviewed in what follows, via the appropriate extension of the JIMWLK equation \cite{Gelis:2010nm}
and its implementation in Langevin form  \cite{Blaizot:2002xy}. (See also \cite{JalilianMarian:2004da,Hentschinski:2005er,Kovner:2006ge,Kovner:2006wr} for previous, closely related, work.) The generalization of this approach to A+A collisions is still an open problem.

\section{\label{sec:qgprod}Quark-gluon production at forward rapidities}

\begin{figure}
\begin{center}
\begin{minipage}[b]{0.49\textwidth}
\begin{center}
\includegraphics[width=0.86\textwidth,angle=0]{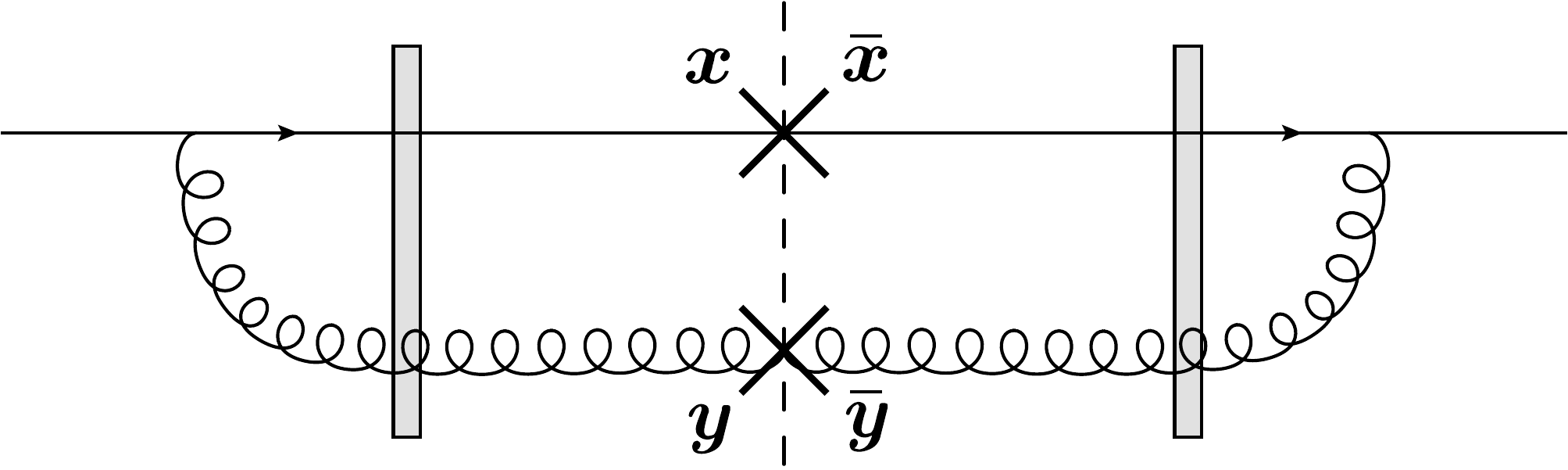}\\(a)
\end{center}
\end{minipage}
\begin{minipage}[b]{0.49\textwidth}
\begin{center}
\includegraphics[width=0.86\textwidth,angle=0]{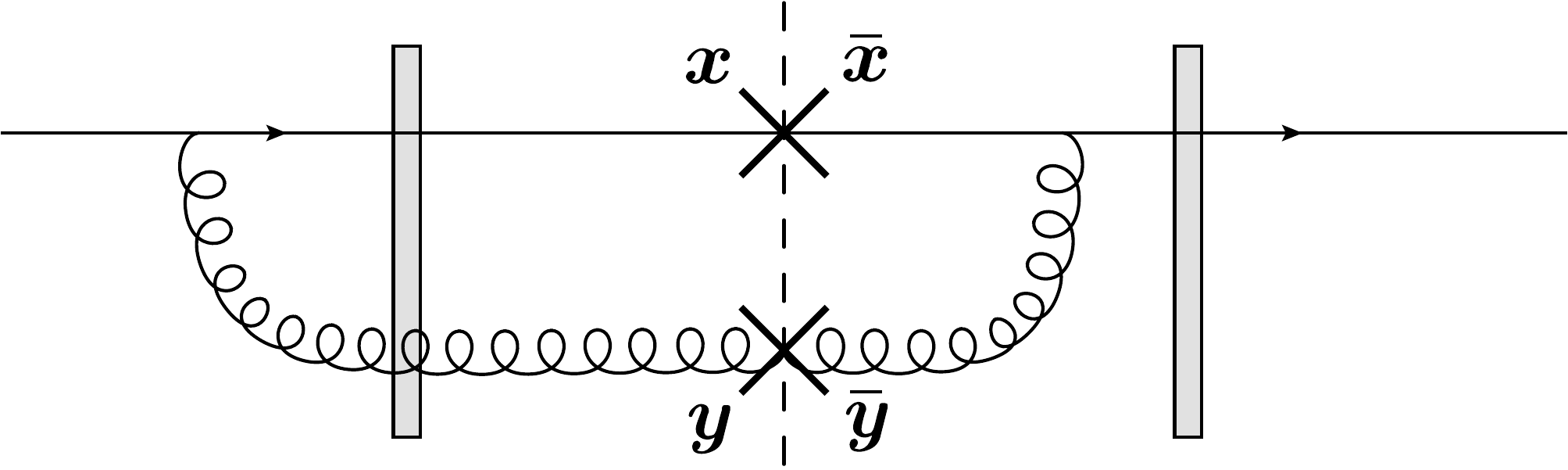}\\(b)
\end{center}
\end{minipage}
\begin{minipage}[b]{0.49\textwidth}
\begin{center}
\includegraphics[width=0.86\textwidth,angle=0]{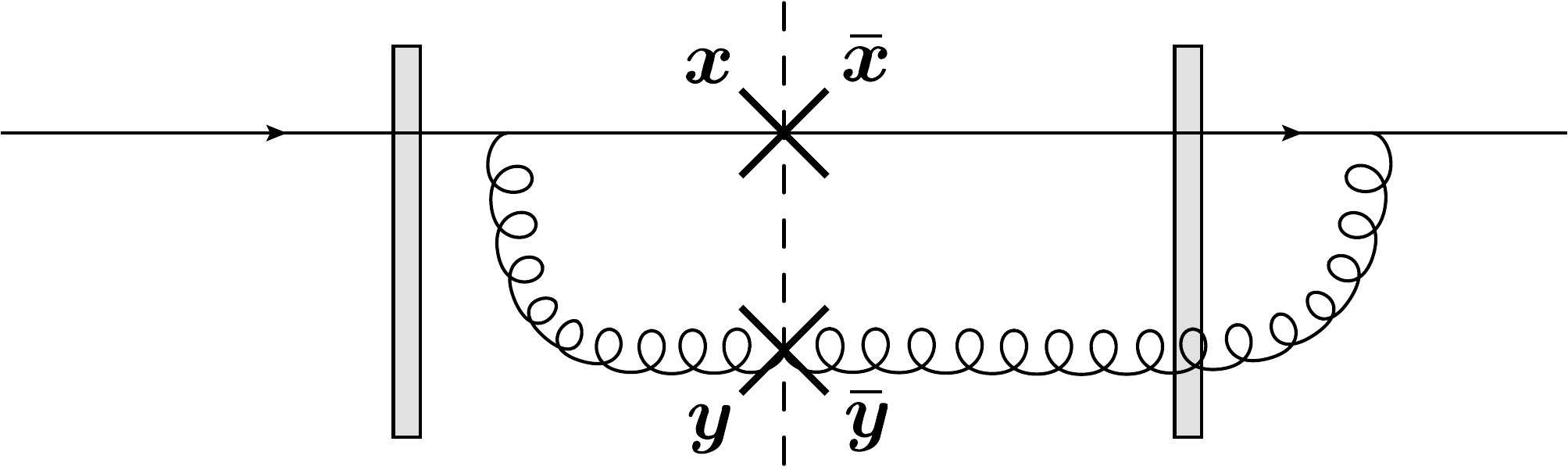}\\(c)
\end{center}
\end{minipage}
\begin{minipage}[b]{0.49\textwidth}
\begin{center}
\includegraphics[width=0.86\textwidth,angle=0]{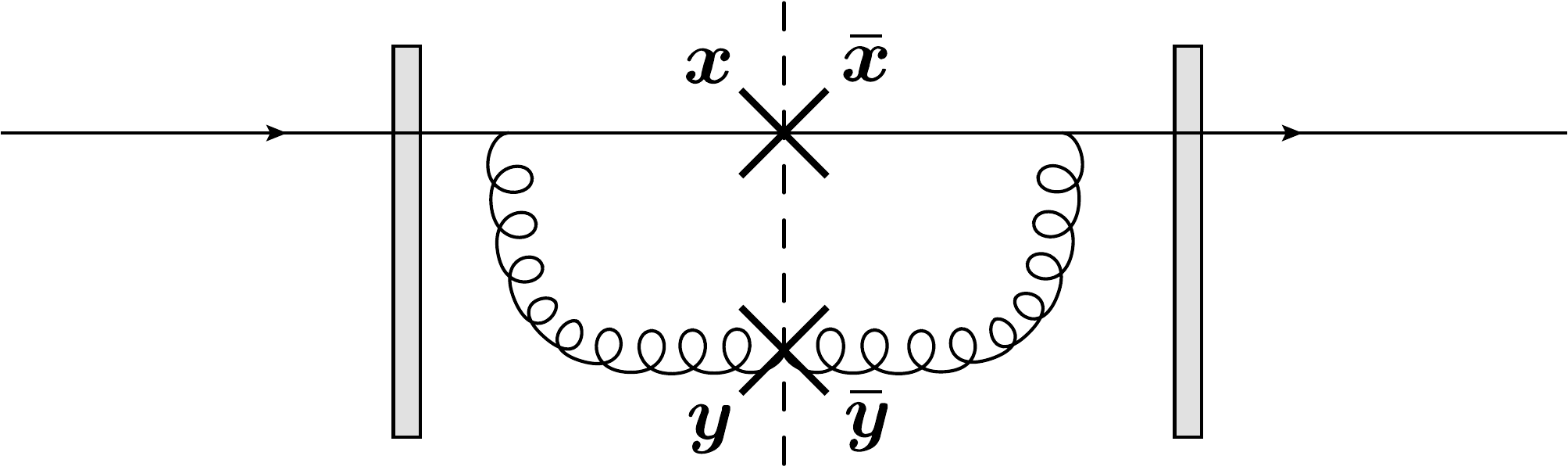}\\(d)
\end{center}
\end{minipage}
\end{center}
\caption{\label{fig:qgprodsame}Diagrams for the production of a quark and a gluon at the same rapidity. A cross stands for each parton produced.}
\end{figure}

The appropriate framework to compute particle production from first principles at high energies is the Color Glass Condensate (CGC), which is an effective theory for gluon saturation endowed with a functional renormalization group equation, the JIMWLK equation \cite{Gelis:2010nm}.  To briefly review this formalism, let us start with a simple problem, which is by now well understood:
the production of a quark--gluon pair, with both partons measured at forward rapidities (that is, in the fragmentation region of the proton), in a high energy p+A collision. We shall describe this process in a frame where most of the energy is carried by the large nucleus. Then the dominant process is that in which a 
quark with a relatively large longitudinal momentum fraction $x$ from the proton (typically,
$x\sim 0.1$) emits a gluon while scattering off the nuclear target (which in this frame
appears as a shockwave, by Lorentz contraction). Both the original quark and the emitted
gluons have large lifetimes as compared to the width of the target, so it is {\em unlikely} that the
emission occurs {\em inside} the nucleus. Rather, the amplitude is controlled by the two processes
where the gluon is emitted either before, or after, the scattering. In the `initial--state
emission' case, the gluon itself can interact with the target. In the `final--state emission' case, the 
gluon does not directly interact, but its emission is indirectly affected by the scattering, 
which modifies the color state of the parent quark. To calculate the cross section, we need to multiply
the direct amplitude (DA)  with the complex conjugate one (CCA), thus obtaining  the four diagrams shown in Fig.~\ref{fig:qgprodsame}.

Due to the high-energy kinematics, one can compute these diagrams in the eikonal approximation:
the transverse coordinates of the partons, $\bm{x}$ for the quark and $\bm{y}$ for the gluon, are preserved by the interaction. The only effect of the collision is a rotation of the parton color state, 
as described by a Wilson line in the appropriate color representation.
However,  since both partons are measured in the final state, one has to 
keep the distinction between the coordinates in the DA and those in the CCA, with the latter denoted by bar, i.e.~$\bar{\bm{x}}$ and $\bar{\bm{y}}$. By taking Fourier transforms with respect to $\bm{x} - \bar{\bm{x}}$ and $\bm{y} - \bar{\bm{y}}$, we can specify the transverse momenta of the produced particles. Let us look, for example, at diagram in Fig.~\ref{fig:qgprodsame}.a which involves
initial--state emissions in both the DA and the CCA. One finds that it is proportional to
 \beq
 \label{qgsame} 
 \alpha_s C_F\, 
 \frac{(\by-\bx)}{(\by-\bx)^2} \cdot
 \frac{(\bby-\bbx)}{(\bby-\bbx)^2}\,
 \frac{2}{N_g}\,
 \Big\langle
 \big(U_{\bby}^{\pd} U_{\by}^\dagger\big)^{ab}
 \rmtr \big[V_{\bx}^\dagger t^b t^a V_{\bbx}^{\pd} \big]
 \Big\rangle_Y, 
 \eeq      
where $N_c$ is the number of colors, $C_F = (N_c^2-1)/2 N_c$, $N_g = N_c^2-1$, $t^a$ is in the fundamental representation and each of the two fractions is the Weizs\"{a}cker-Williams kernel for the emission of a soft gluon by a point-like color source. We use different notations for the quark ($V,\,V^{\dagger}$) and gluon  ($U,\,U^{\dagger}$) Wilson lines. For instance,
 \beq
 \label{uwl}
 U^{\dagger}_{\bx} = {\rm T}
  \exp \left[ \rmi g \int \dif x^+ 
  A_a^{-}(x^+,\bx) T^a \right]
\eeq
with $T^a$ in the adjoint. Here we have introduced $x^+ = (t+ x^3)/\sqrt{2}$ and assumed that the projectile proton moves along the   
positive $x^3$-direction and thus it interacts with the $A^-$ component of the color field of the nucleus. The average in \eqn{qgsame} refers to the target wavefunction and must be computed with
the CGC weight function, which describes the distribution of the  random color field $A^-$ 
inside the nucleus (see \eqn{save} below). This distribution depends upon the rapidity 
$Y$ at which one probes the nuclear 
gluon distribution, which is fairly large (say, $Y\gtrsim 3$) for forward kinematics. The
evolution of the Wilson line correlators with $Y$ is described by the JIMWLK  equation (or, equivalently, by the Balitsky hierarchy)  \cite{Gelis:2010nm}, to which we shall return in a while. At large $N_c$,
the Wilson line correlator in \eqn{qgsame} --- and, more generally, all the similar correlators 
that are probed by multi-particle production --- can be written in terms of color dipoles and 
quadrupoles \cite{JalilianMarian:2004da,Kovner:2006wr,Dominguez:2012ad}.

\section{Forward--central production and the lack of factorization}   

\begin{figure}
\begin{center}
\begin{minipage}[b]{0.33\textwidth}
\begin{center}
\includegraphics[width=0.99\textwidth,angle=0]{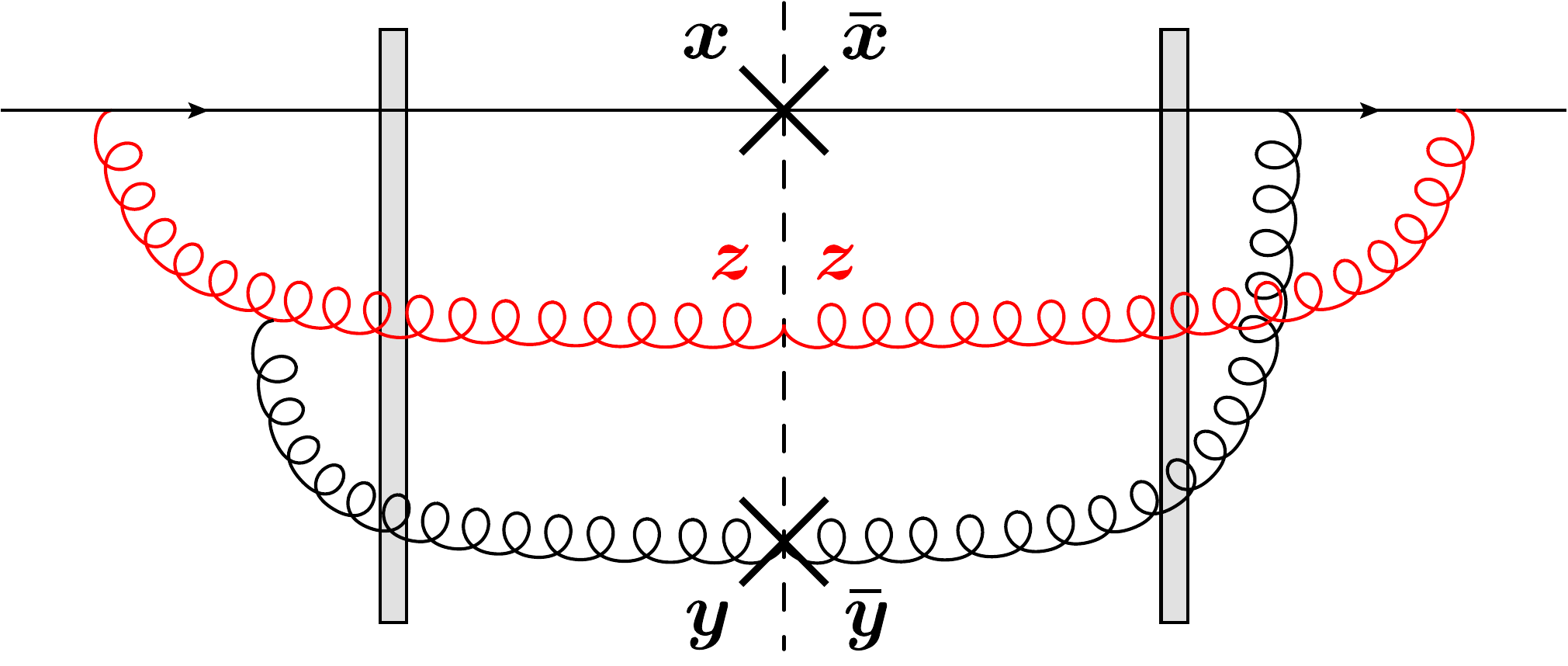}\\(a)
\end{center}
\end{minipage}
\begin{minipage}[b]{0.33\textwidth}
\begin{center}
\includegraphics[width=0.99\textwidth,angle=0]{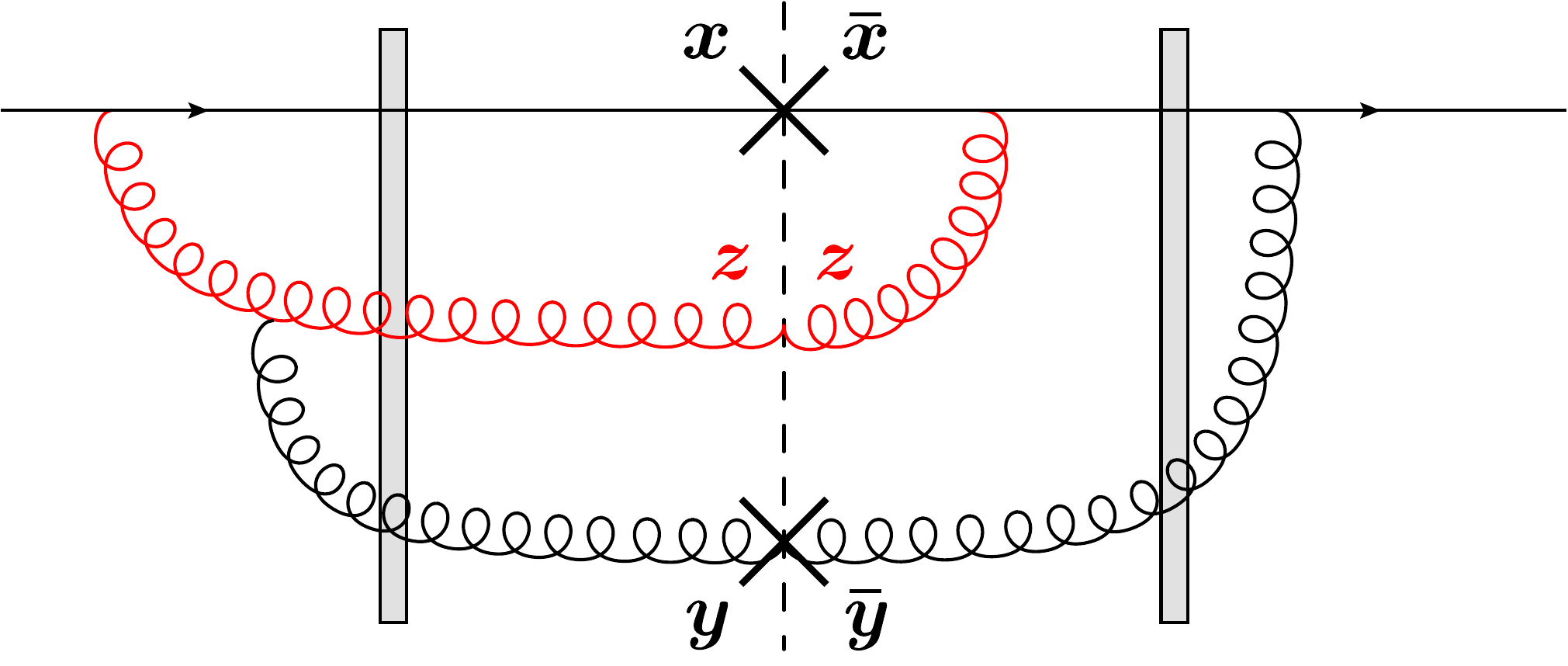}\\(b)
\end{center}
\end{minipage}
\begin{minipage}[b]{0.33\textwidth}
\begin{center}
\includegraphics[width=0.99\textwidth,angle=0]{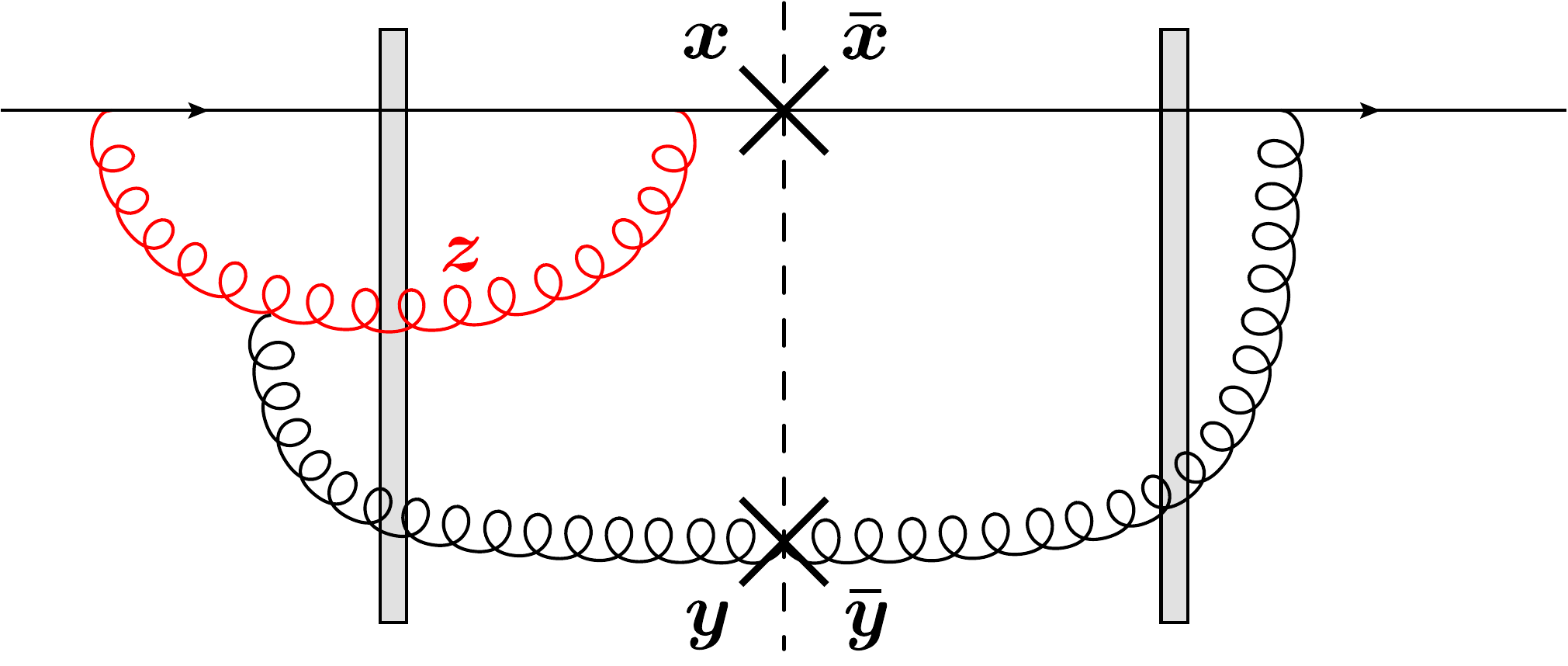}\\(c)
\end{center}
\end{minipage}
\end{center}
\caption{\label{fig:qgevol}Diagrams for the production of a quark and a gluon accompanied by an intermediate unresolved gluon.}
\end{figure}

In the previous discussion, we implicitly assumed that the rapidity difference $\Delta Y$
between the two produced partons was relatively small, $\Delta Y\ll 1/\alpha_s$, so in particular
they both probe the nuclear wavefunction at the same rapidity $Y$. From now on, we turn
to the actual problem of interest for us here, which is the case of a large rapidity separation 
$\Delta Y \gtrsim 1/\alpha_s$. Then one needs to take into account the `high-energy evolution', that is, the emission of unresolved gluons, between the two measured partons. By appropriately choosing
the frame, one can associate such additional emissions with either the projectile, or the target,
 and for the time being it is convenient to adopt the projectile viewpoint.
Fig.~\ref{fig:qgevol} exhibits some of the diagrams where only one intermediate gluon, at transverse coordinate $\bz$, has been emitted by the quark. This includes both `real' emissions, where the
`evolution' gluon is produced in the final state (albeit it is not measured) --- the corresponding
line crosses the cut, as shown in Figs.~\ref{fig:qgevol}.a and \ref{fig:qgevol}.b --- 
and `virtual' emissions, in which the gluon is emitted and reabsorbed
on the same side of the cut, cf. Fig.~\ref{fig:qgevol}.c.

The first process, in Fig.~\ref{fig:qgevol}.a, is simpler in the sense that the interactions of the
 `evolution' gluon with the nucleus cancel between the DA and the CCA by unitarity, $U^{\pd}_{\bz} U^{\dagger}_{\bz} =1$. This is possible because this gluon is not measured (its
transverse position is the same in the DA and the CCA) and because the evolution occurs in the
initial state (before the collision), in both the DA and the CCA. So, in principle, it is possible
to view this emission as part of the quark wavefunction prior to the collision and thus factorize
it out from the interaction. This strategy is at the basis of standard factorization schemes for
particle production, like the collinear or the $\bk_{\perp}$--factorization. However,
for the problem at hand, the situation is complicated by the existence of single--cross diagrams, like
those in Figs.~\ref{fig:qgevol}.b and \ref{fig:qgevol}.c, which are associated with final--state
evolution. In such diagrams, the evolution gluon crosses the shockwave only once, hence
its interaction (represented by the Wilson line $U_{\bz}^\dagger$ in the DA)
survives in the final result. Thus, not only the measured partons, but also the unresolved ones, know about the nucleus, and the factorization is bound to fail in general. 

There are two special
cases though where the $\bk_{\perp}$--factorization is recovered. One is the case where the
original quark is not measured, i.e.~when we set $\bbx=\bx$. Then, the diagrams in Figs.~\ref{fig:qgevol}.b and \ref{fig:qgevol}.c differ just be a sign, since the latter can be obtained from the former by moving 
one endpoint of the evolution gluon from the CCA to the DA. Hence, these two diagrams sum up to zero. This observation is generic: if one measures just one particle, or a set of particles with similar
rapidities, then the only relevant evolution on the projectile side is the BFKL (initial--state)
evolution of the respective wavefunction. The other case where $\bk_{\perp}$--factorization is
resuscitated is for a dilute target, so that multiple scattering can be neglected. Then the scattering
of the evolution gluons can be neglected (as a higher--order effect) and factorization applies
even for relatively large rapidity separations.

In what follows, however, we shall not be interested in such special limits, but rather in
the general case for which there is no factorization. This includes interesting physical situations
like the production of a pair of particles at forward--central rapidities and the `ridge' effect in
p+A collisions. In all such cases, one needs to be able to follow the BFKL evolution of a
dilute projectile in the presence of a strong background field  --- the color field of the target.

\section{JIMWLK evolution for scattering amplitudes}

Before we immerse ourselves in particle production, let us consider a related but simpler problem where the BFKL evolution in a strong background field is well understood. This is the B--JIMWLK evolution of the {\em scattering amplitudes} for the
collision between a dilute projectile (proton) and a dense target (nucleus)  \cite{Gelis:2010nm,Balitsky:2001gj}. 
At high
energy, these amplitudes are built with gauge--invariant products of Wilson lines, like
that in \eqn{qgsame}, which describe the elastic scattering of the projectile partons
off the strong target color field. Inelastic effects are generated when taking into
account the target correlations, i.e.~when averaging 
this product of Wilson lines over all the possible configurations of the target field with the CGC weight function.
Via unitarity, the average amplitudes enter the calculation of the total cross--sections and of particle production at similar rapidities, as in the case of 
quark--gluon production in Sect.~\ref{sec:qgprod}.

\begin{figure}
\begin{center}
\begin{minipage}[b]{0.33\textwidth}
\begin{center}
\includegraphics[width=0.95\textwidth,angle=0]{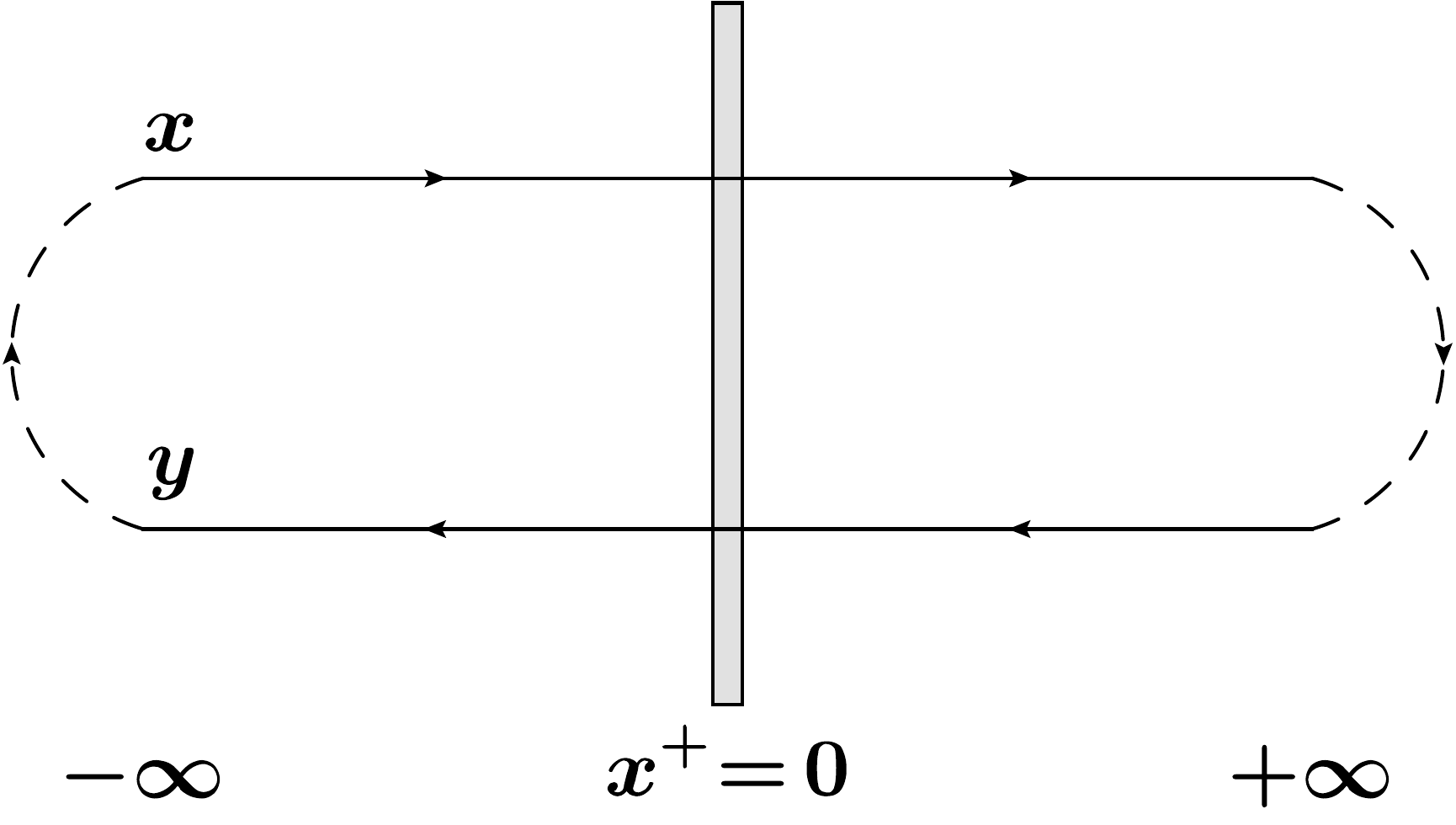}\\(a)
\end{center}
\end{minipage}
\begin{minipage}[b]{0.33\textwidth}
\begin{center}
\includegraphics[width=0.95\textwidth,angle=0]{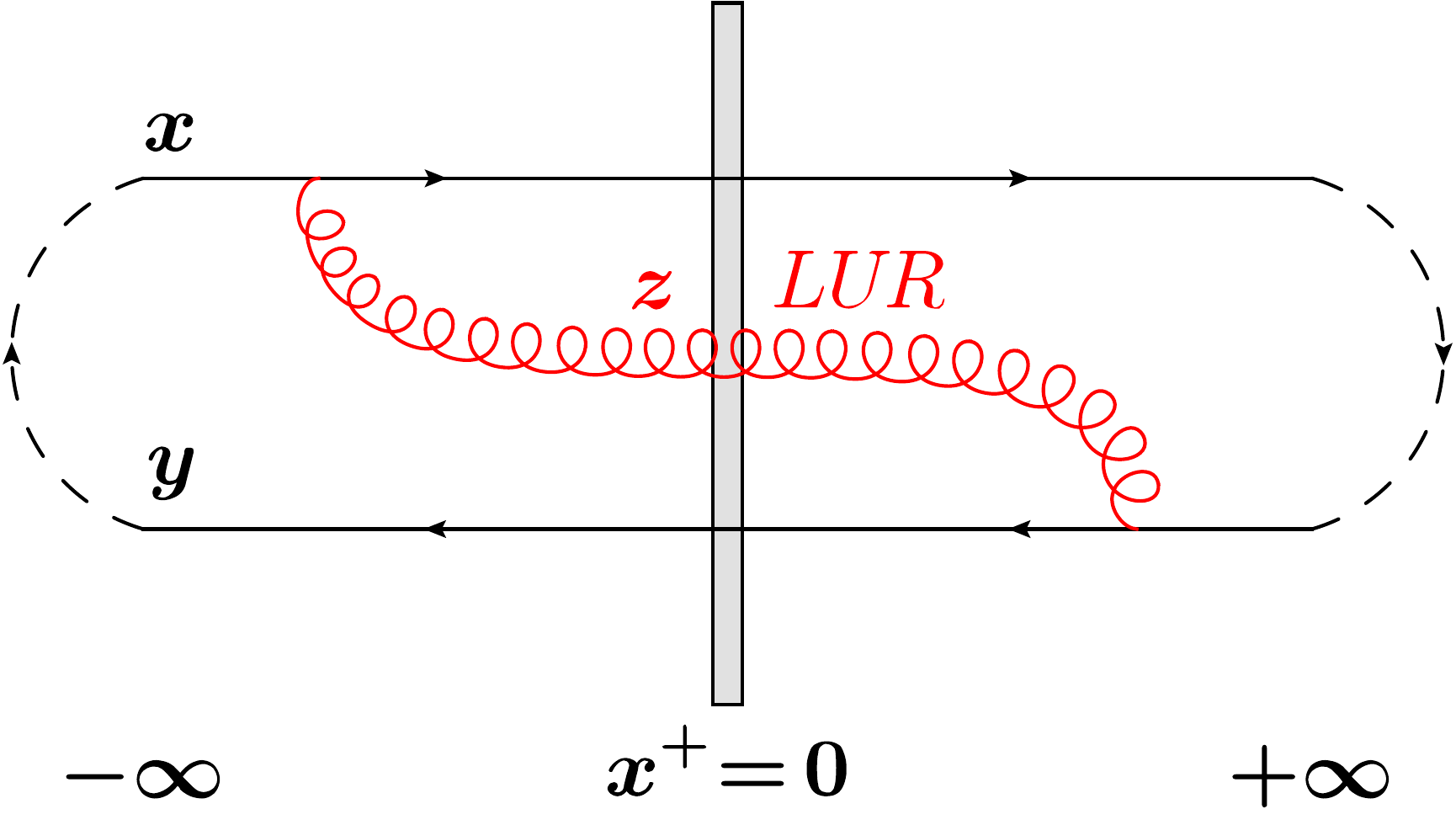}\\(b)
\end{center}
\end{minipage}
\begin{minipage}[b]{0.33\textwidth}
\begin{center}
\includegraphics[width=0.95\textwidth,angle=0]{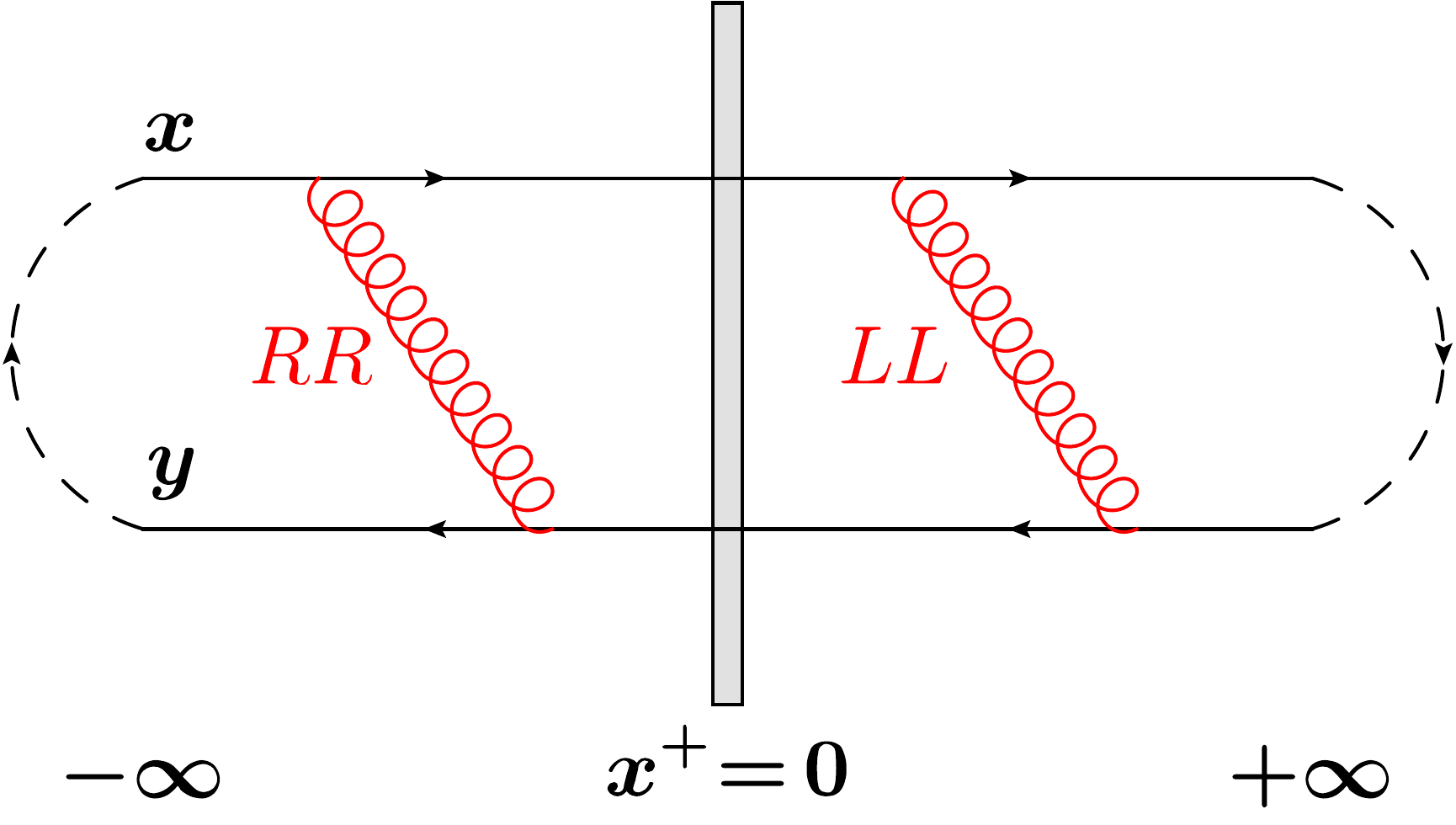}\\(c)
\end{center}
\end{minipage}
\end{center}
\caption{\label{fig:dipevol}Diagrams for the dipole $S$-matrix and its evolution.}
\end{figure}

For definiteness, we consider the simplest gauge--invariant projectile: a quark--antiquark
dipole. The average value of the elastic $S$--matrix
for dipole--nucleus scattering reads (see Fig.~\ref{fig:dipevol}.a)
 \beq
 \label{save}
 \big\langle S_{\bx\by}\big\rangle_Y = 
 \int \mcal{D}A^- \, W_Y[A^-]\,
 \frac{1}{N_c}\,
 \rmtr \big[V_{\by} V_{\bx}^{\dagger} \big]\,,
 \eeq
where $W_Y[A^-]$  is the CGC weight function --- the probability to find a given configuration 
for the color field $A^-_a$ in the target ---, which plays the role of
the {\em target wavefunction squared}. \eqn{save} is written in a frame where 
the total rapidity difference $Y$ is carried by the nucleus.

Let us consider a rapidity increment $Y\to Y+\rmd Y$. The additional rapidity $\rmd Y$ 
can be given either to the projectile, or to the target, resulting in different pictures for the evolution, 
but with the same result for the change $\rmd\big\langle S_{\bx\by}\big\rangle_Y$ in the average
$S$--matrix. Consider first the evolution of the projectile (the dipole). With probability $\alpha_s
\rmd Y$, this can undergo a gluon emission and absorption, which can be either fully in
the initial state, or in the final state, or mixed, with examples shown in the last two diagrams in Fig.~\ref{fig:dipevol}. (Clearly, there can be also diagrams in which both ends of the gluon are attached to the same quark.) Therefore, as clear from the diagram in Fig.~\ref{fig:dipevol}.b, this gluon can also interact with the strong target field. Such an evolution is described by the JIMWLK Hamiltonian 
\cite{Gelis:2010nm}, more precisely
 \beq
 \label{dsdy}
 \frac{\del}{\del Y}\, S_{\bx\by} = H_{\rm JIMWLK}\, S_{\bx\by},  
 \eeq
which gives the change in the scattering operator for a fixed target field.
The JIMWLK Hamiltonian reads
 \beq
 \label{hjimwlk}
 H_{\rm JIMWLK} = 
 \frac{1}{8\pi^3}
 \int \dif^2\bu\, \dif^2\bw\, \dif^2\bz\,
 \frac{\bz - \bu}{(\bz-\bu)^2} \cdot
 \frac{\bz - \bw}{(\bz-\bw)^2}\,
 \big[L^a_{\bu} - U^{\dagger ab}_{\bz} R^b_{\bu} \big]
 \big[L^a_{\bw} - U^{\dagger ac}_{\bz} R^c_{\bw} \big],
 \eeq
where the ``Right'' and ``Left'' Lie derivatives correspond to gluon emissions (or absorptions) before and after the scattering as shown in Figs.~\ref{fig:dipevol}.b and \ref{fig:dipevol}.c, and are given by
 \beq
 \label{rl}
 R^a_{\bu} U^{\dagger}_{\bx} = 
 \rmi g \delta_{\bu\bx} U^{\dagger}_{\bx} T^a,
 \qquad
 L^a_{\bu} U^{\dagger}_{\bx} = 
 \rmi g \delta_{\bu\bx} T^a U^{\dagger}_{\bx} 
 = U^{\dagger ab}_{\bu} R^b_{\bu} U^{\dagger}_{\bx}.
 \eeq  
The Lie derivatives are truly operators measuring the color charge density in the projectile.

After averaging the dipole equation \eqref{dsdy} over the target and
taking the large--$N_c$ limit, one obtains a closed equation, the Balitsky--Kovchegov equation, 
whose solution has been studied at length \cite{Gelis:2010nm}. 
However, at finite $N_c$ one needs to consider an infinite hierarchy of equations,
the B--JIMWLK hierarchy, which is hopelessly complicated.
Furthermore, even at large $N_c$ but for a more complicated projectile, like the quadrupole 
which enters di--hadron production at  forward rapidities (cf. \eqn{qgsame}),
the associated equations are extremely cumbersome (even though a tractable approximation
scheme has been recently devised in \cite{Kovchegov:2008mk,Dumitru:2011vk,Iancu:2011ns,Iancu:2011nj,Alvioli:2012ba}). A more efficient method to address such problems via numerical 
simulations will be described in the next section.

\section{\label{sec:lang} A Langevin approach to JIMWLK evolution}

\begin{figure}
\begin{center}
\begin{minipage}[b]{0.4\textwidth}
\begin{center}
\includegraphics[width=0.78\textwidth,angle=0]{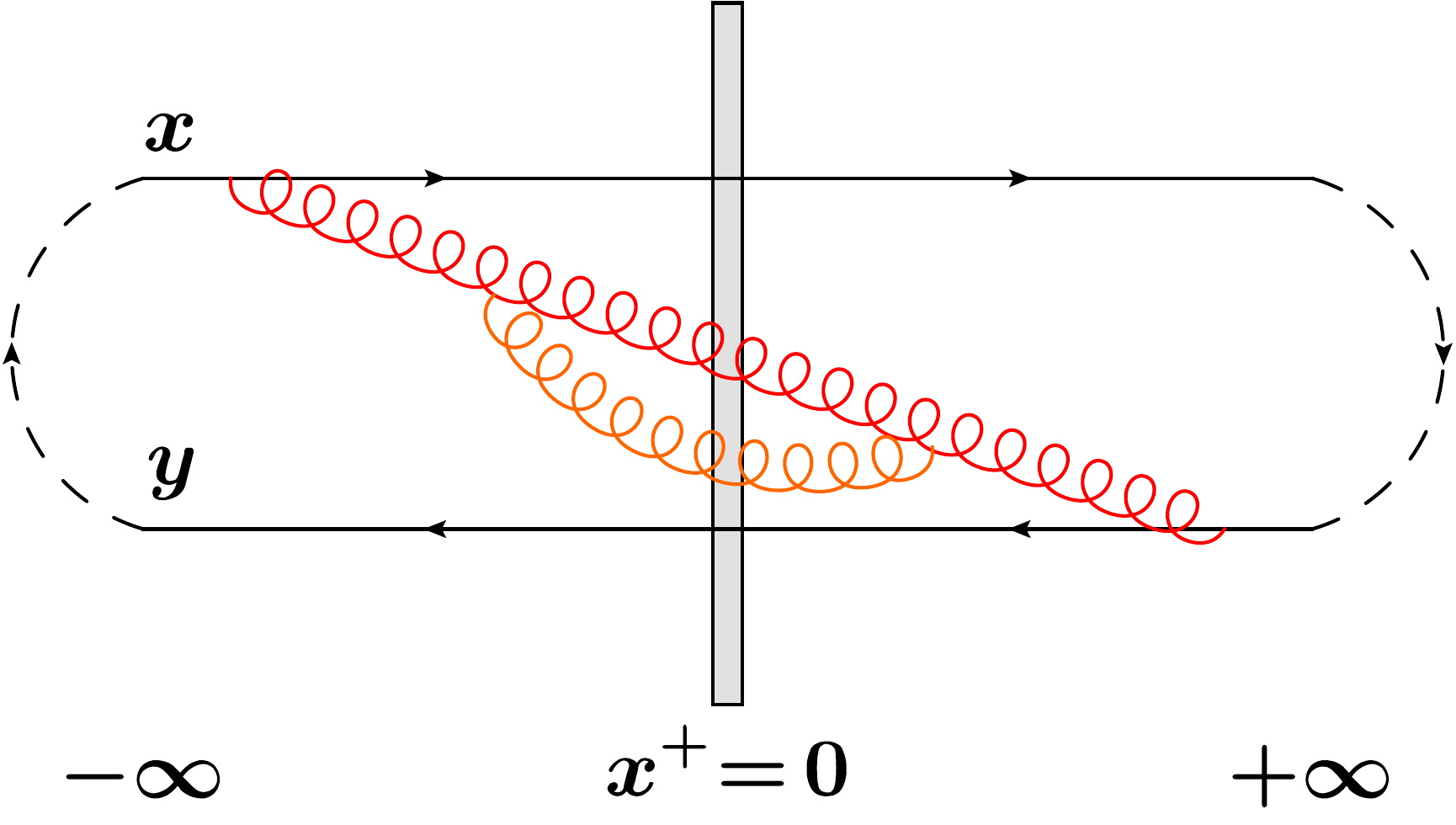}\\(a)
\end{center}
\end{minipage}
\begin{minipage}[b]{0.4\textwidth}
\begin{center}
\includegraphics[width=0.78\textwidth,angle=0]{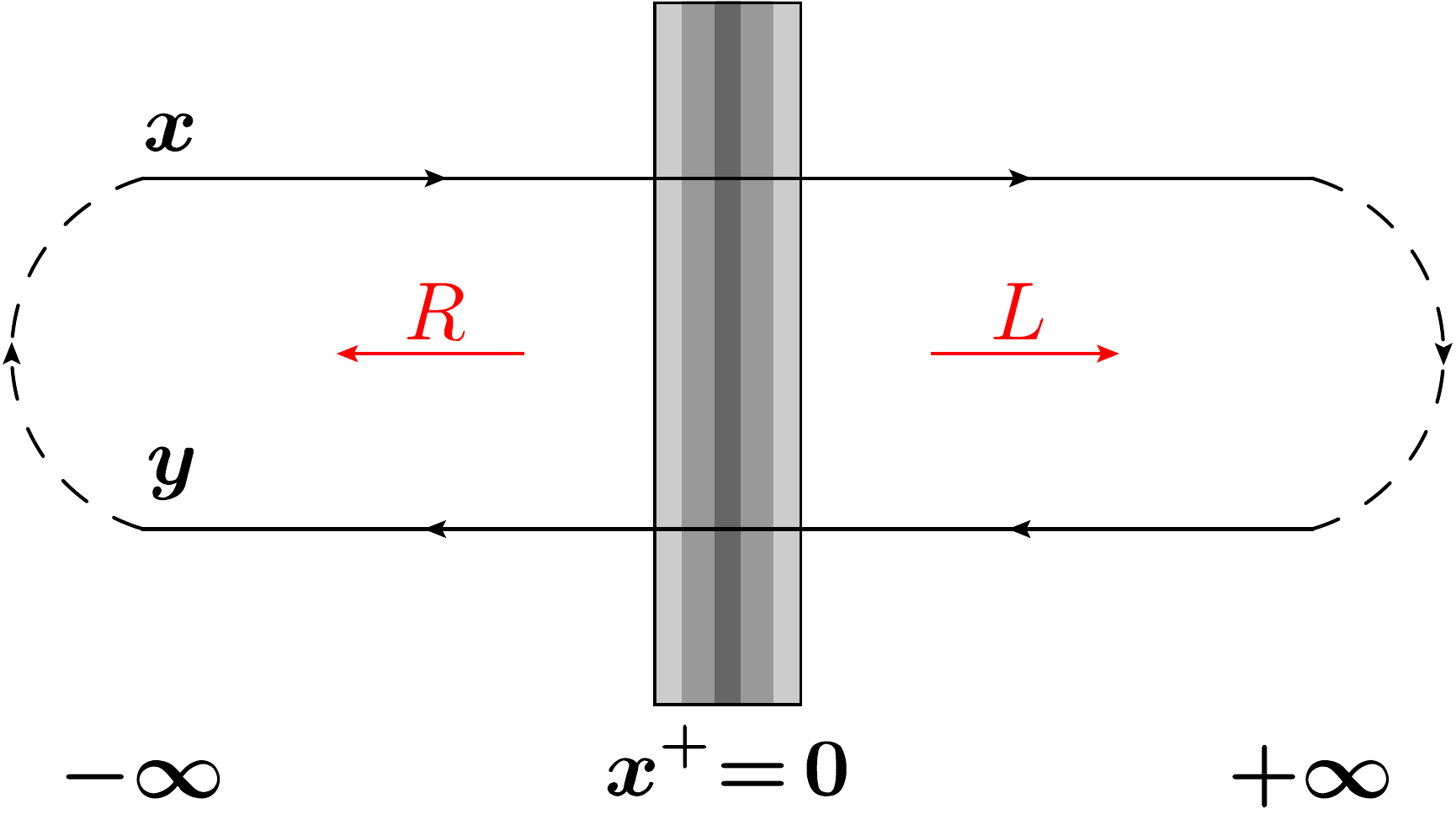}\\(b)
\end{center}
\end{minipage}
\end{center}
\caption{\label{fig:projtar}Diagrams for (a) projectile and (b) target evolution.}
\end{figure}

This alternative method privileges the viewpoint of {\em target evolution}, which is the original
JIMWLK approach. Namely, the JIMWLK equation is truly a (functional) evolution equation 
for the target weight function $W_Y[U]$ introduced in \eqn{save}. 
From this perspective, the projectile remains `bare' but the target becomes `fatter' in the $x^+$-direction (the longitudinal direction for the left--moving nucleus), due to the additional
color charges generated by the quantum evolution (see Fig.~\ref{fig:projtar}). 
One step of target evolution consists in the emission of a gluon which is `softer' --- in the sense
of having a smaller $k^-$, or a larger longitudinal wavelength $\Delta x^+=1/k^-$ --- than all of its ancestors. 
This adds a new layer to the target field at larger values of $|x^+|$
and modifies the Wilson lines describing the scattering of dilute a projectile. This new layer
is a result of a quantum calculation, so is itself random. Accordingly, this evolution is naturally
stochastic and can conveniently be formulated as a Langevin equation \cite{Blaizot:2002xy}.

To be more specific, let us introduce a discretization of the rapidity interval according to 
$Y=N \epsilon$. Then the JIMWLK evolution is equivalent to the following
recurrence formula for the Wilson lines:
  \beq
 \label{un}
 U^{\dagger}_{n,\bx}
 =\rme^{\rmi \epsilon g \alpha^L_{n,\bx}}
 U^{\dagger}_{n-1,\bx}
 \rme^{-\rmi \epsilon g \alpha^R_{n,\bx}}
 \eeq
with the `left' and `right' gauge fields determined by
 \beq
 \label{alar}
 \alpha^L_{n,\bx} = 
 \frac{1}{\sqrt{4\pi^3}}
 \int \dif^2\bz\, 
 \frac{\bx^i - \bz^i}{(\bx-\bz)^2}\, 
 \nu^{ia}_{n,\bz} T^a ,
 \qquad
 \alpha^R_{n,\bx} = 
 \frac{1}{\sqrt{4\pi^3}}
 \int \dif^2\bz\, 
 \frac{\bx^i - \bz^i}{(\bx-\bz)^2}\,
 U^{ab}_{n-1,\bz} 
 \nu^{ib}_{n,\bz} T^a\,,
 \eeq
with $\nu^{ia}_{n,\bz}$ a noise term which accounts for the color charge density and the polarization of the evolution gluons radiated in the step under consideration. This noise is Gaussian and `white',
 in the sense that 
  \beq
 \label{nu}
 \big\langle
 \nu^{ia}_{m,\bx} \nu^{jb}_{n,\by}
 \big\rangle
 = \frac{1}{\epsilon}\,
 \delta^{ij} \delta^{ab} 
 \delta_{mn} \delta_{\bx\by}\,.
 \eeq
Note that the Wilson $U$ in the above expression for $\alpha^R$ describes the
interactions between the newly emitted gluon and the preexisting 
color charges and represents the origin of the BFKL cascades in the present approach. 
To fully specify this stochastic process, one also needs an initial condition, i.e.~a choice for
the Wilson line $U^{\dagger}_{0,\bx}$ in the first step.
A standard procedure is to randomly select this Wilson line
according to the McLerran--Venugopalan model \cite{Gelis:2010nm}. 
Within this Langevin formalism, the expectation values at rapidity $Y$ are obtained as,
e.g. 
 \beq
 \label{savelang}
 \big\langle
 S_{\bx\by}
 \big\rangle_Y 
 =\frac{1}{N_c}
 \Big\langle 
 \rmtr \big[ V^\pd_{N,\by}
 V^{\dagger}_{N,\bx} \big]
 \Big\rangle_{\nu}\,,
 \eeq
 where the brackets refer to the average over the noise at all the intermediate steps
 $n\le N$, according to \eqref{nu}. As compared to the B--JIMWLK hierarchy,
 this stochastic procedure is better suited for numerical implementations and has 
 indeed allowed for explicit, numerical, solutions \cite{Rummukainen:2003ns,Dumitru:2011vk}.
 
\section{JIMWLK evolution for multi--particle production}

\begin{figure}
\begin{center}
\begin{minipage}[b]{0.49\textwidth}
\begin{center}
\includegraphics[width=0.95\textwidth,angle=0]{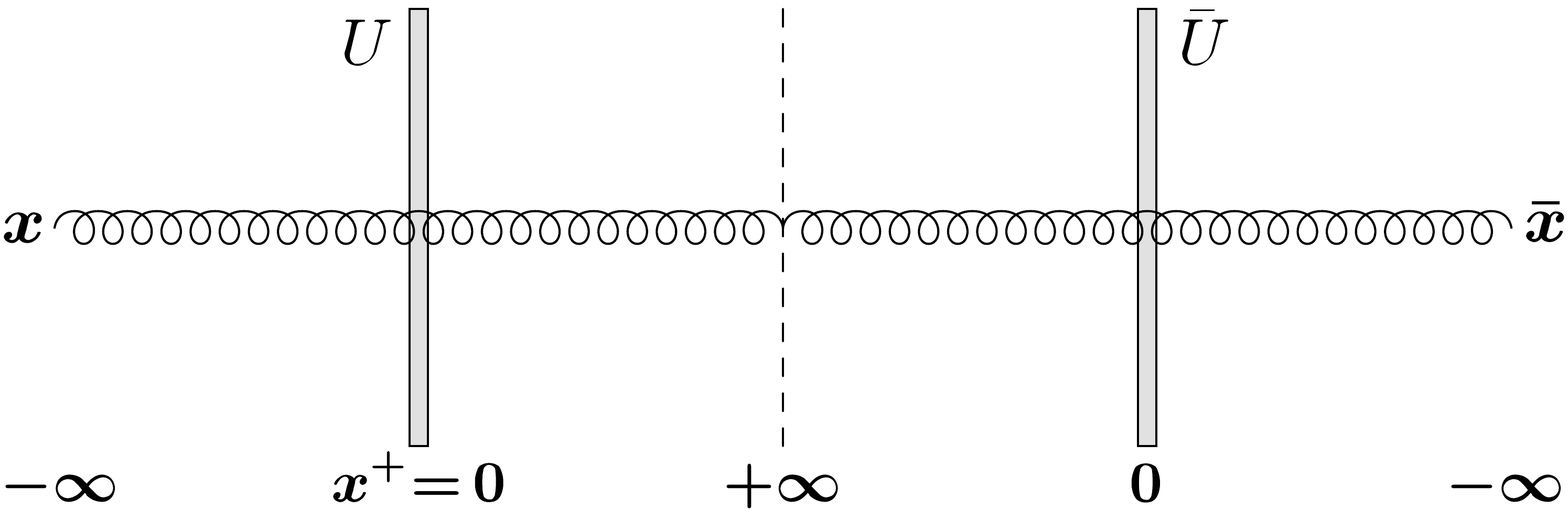}\\(a)
\end{center}
\end{minipage}
\begin{minipage}[b]{0.49\textwidth}
\begin{center}
\includegraphics[width=0.95\textwidth,angle=0]{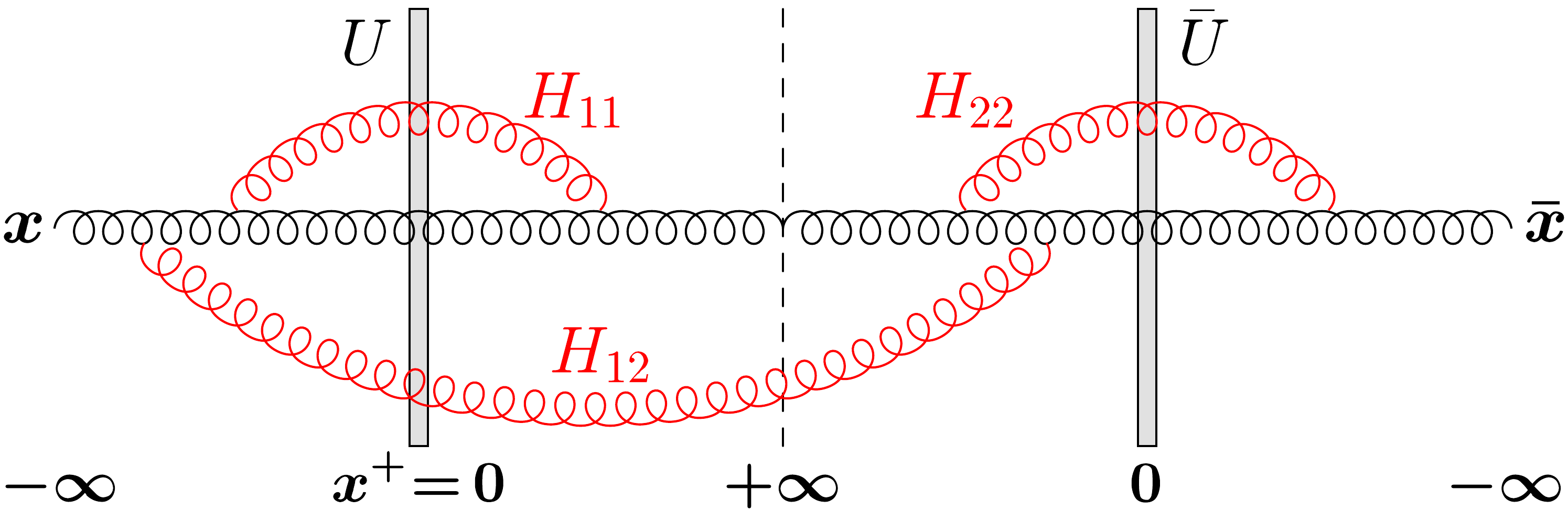}\\(b)
\end{center}
\end{minipage}
\end{center}
\vspace*{-0.1cm}
\caption{\label{fig:genfun}Diagrams for (a) the generating functional of a bare gluon and (b) its one step evolution.}
\end{figure}

We now return to multi--particle production with large rapidity separation(s) 
in p+A collisions, and show how it can be computed within a suitable generalization of the CGC effective theory \cite{Hentschinski:2005er,Kovner:2006ge,Kovner:2006wr,Iancu:2013uva}. We need to extend our formalism from {\em amplitudes} to {\em cross--sections} and to this end a convenient tool is the {\em generating functional} for soft gluon emissions,
resolved or not, out of a dilute partonic system. To understand this concept, notice that the
functional $S_{\bx\by}[A^-] = ({1}/{N_c})\rmtr \big[V_{\by} V_{\bx}^{\dagger} \big]$ representing the dipole
$S$--matrix can also be seen as the {\em wavefunction} of a bare dipole in the $A^-$--representation
and in the eikonal approximation. The generating functional can be similarly viewed as a gauge--invariant
{\em transition amplitude}
from a given configuration $\{\bx, \by, A^-\}$ in the DA to another configuration
$\{\bbx, \bby, \bar A^-\}$ in the CCA. For more clarity, consider a simpler projectile, e.g.~a gluon. The generating functional 
for a {\em bare} gluon reads
 \beq
 \label{s12}
 S_{12}(\bx\bbx) =
 \frac{1}{N_g} \rmTr \big[ \bar{U}_{\bbx} 
 U^{\dagger}_{\bx}\big],
 \eeq
where $\bx$ and $\bbx$ are the transverse coordinates of the gluon in the DA and the CCA respectively,
and $U$ and $\bar{U}$ are the corresponding adjoint Wilson lines, which are a priori independent 
from each other.
Hence, $S_{12}(\bx\bbx) $ should be viewed as a functional of $U$ and $\bar{U}$. The color trace
in \eqn{s12} has been generated by the sum (average) over the final (initial) gluon color indices. A
diagrammatic representation for this equation is given in Fig.~\ref{fig:genfun}.a. 

It is clear in this example, that there is a doubling of the degrees of freedom: a physical gluon is described by a mathematical dipole, a physical dipole will be described by a mathematical quadrupole, and so on.  This is 
reminiscent of the Schwinger--Keldysh formalism and it occurs for the same basic reason:
it allows us to independently emit  gluons from the DA and the CCA, and thus build
a cross--section. As before, the soft gluon emissions are generated by Lie derivatives acting on the
Wilson lines, but now one has to distinguish between `barred' and `unbarred' Lie derivatives.

For instance, to produce a gluon with transverse momentum $\bk$ from a dilute projectile, one should
act on the corresponding generating functional with the `production Hamiltonian'
 \beq
 \label{hprod}
 H_{\rm prod}(\bk) = 
 \frac{1}{4 \pi^3}
 \int \dif^2\bu\,\dif^2\bw\,\dif^2\by\,\dif^2\bby\,
 \rme^{-\rmi \bk \cdot(\by - \bby)}\,
 \frac{\by - \bu}{(\by-\bu)^2} \cdot
 \frac{\bby - \bw}{(\bby-\bw)^2}\,
 \big[L^a_{\bu} - U^{\dagger ab}_{\by} R^b_{\bu} \big]
 \big[\bar{L}^a_{\bw} - \bar{U}^{\dagger ac}_{\bby} \bar{R}^c_{\bw} \big].
 \eeq
Obviously the above shows many similarities with the JIMWLK Hamiltonian \eqref{hjimwlk}, but 
also some
 differences: there are now two types of Wilson lines and Left/Right Lie derivatives, `barred' and `unbarred',
and the gluon is radiated at different transverse positions in the DA and in the CCA, since it must
be produced with a given momentum $\bk$. In particular, if the original projectile is a bare gluon and the two gluons have similar rapidities 
$Y_p$ and $Y_k$, one has     
 \beq
 \label{sigma2g}
 \frac{\dif \sigma_{2g}}{\dif Y_{p} \dif^2\bp \,\dif Y_{k} \dif^2\bk}
 = \frac{1}{(2\pi)^4}\,
 \int \dif^2 \bx \, \dif^2 \bbx\,
 \rme^{-\rmi \bp \cdot(\bx - \bbx)}
 \Big\langle H_{\rm prod}(\bk)\, S_{12}(\bx\bbx)\big|_{\bar{U}=U} 
 \Big\rangle_Y.
 \eeq 
The action of $H_{\rm prod}(\bk)$ on $S_{12}(\bx\bbx) $ generates 
four diagrams similar to the ones for quark--gluon production in Fig.~\ref{fig:qgprodsame}. After producing the gluon, the Wilson lines in the DA and the CCA are identified with
each other, e.g.~$\bar{U}_{\bbx} \to U_{\bbx}$, and with the {\em physical} Wilson lines generated
by the target color field. This field is finally averaged out in line with \eqn{save}.

\begin{figure}
\begin{center}
\includegraphics[width=0.6\textwidth,angle=0]{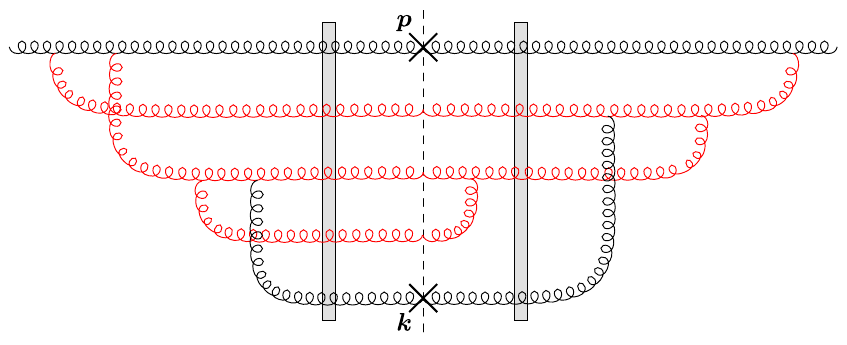}
\end{center}
\caption{\label{fig:digluon}Typical diagram for the production of two different gluon at different rapidities.}
\end{figure}

Now we wish to increase the rapidity separation $\Delta Y=Y_p-Y_k$ between the
two produced gluons, up to rather large values $\Delta Y \gtrsim 1/\alpha_s$, where the high energy evolution becomes important. 
Diagrammatically, we have to systematically generate and calculate all the Feynman graphs like that in Fig.~\ref{fig:digluon},
where the number of intermediate `evolution' gluons can be arbitrary.
To that aim, we need to evolve the generating functional of the original gluon by including
all the unresolved, soft, gluon emissions within the interval $\Delta Y$. This evolution is governed
by an equation similar to \eqn{dsdy}, but with the JIMWLK Hamiltonian replaced by its extension
to the Schwinger--Keldysh time contour, and reads \cite{Hentschinski:2005er,Kovner:2006ge,Kovner:2006wr}
 \beq
 \label{hevol}
 H_{\rm evol}[U,\bar{U}] =
 H_{11}[U] 
 +H_{22}[\bar{U}]
 + 2 H_{12}[U,\bar{U}].
 \eeq 
$H_{11}[U]$ is the JIMWLK Hamiltonian in \eqn{hjimwlk}, $H_{22}[\bar{U}]$ is obtained by barring all the Wilson lines and derivatives in \eqn{hjimwlk}, while $H_{12}[U,\bar{U}]$ is obtained by barring only the second square bracket in that equation. In Fig.~\ref{fig:genfun}.b we show examples of gluon emissions
generated by $H_{\rm evol}$ when acting on the bare gluon generating functional.

Starting with \eqn{hevol}, it
is possible to construct generalized B--JIMWLK equations for the generating functionals, or directly for the cross sections for multi--parton production. These equations appear to be hopelessly complicated even at 
large--$N_c$ \cite{JalilianMarian:2004da,Kovner:2006wr,Iancu:2013uva} and a Langevin
reformulation appears as the only convenient approach in practice \cite{Iancu:2013uva}. 

\section{Langevin reformulation for multi--particle production}

Once again, the  Langevin formulation is most convenient understood from the viewpoint of target evolution. 
With reference to Fig.~\ref{fig:digluon}, let us call $Y_A$ the rapidity of the slowest produced gluon 
(with transverse momentum $\bk$) and $Y= Y_A + \Delta Y$ that of the fastest  one
(with transverse momentum $\bp$).  We shall now construct the intermediate evolution over the rapidity 
interval $\Delta Y$ {\em upwards}, i.e.~from $Y_A$ up to $Y$,
starting with generic Wilson lines $U^{\dagger}_A$ and $\bar{U}^{\dagger}_A$ at $Y_A$. Besides
the Langevin process in the DA as discussed in Sect.~\ref{sec:lang}, we also need a similar
process for the CCA, namely
 \beq
 \label{unbar}
 \bar{U}^{\dagger}_{n,\bx}
 =\rme^{\rmi \epsilon g \bar{\alpha}^L_{n,\bx}}
 \bar{U}^{\dagger}_{n-1,\bx}
 \rme^{-\rmi \epsilon g \bar{\alpha}^R_{n,\bx}}\,,
 \eeq
where the barred left and right fields involve the {\em same} noise term $\nu^{ia}_{n,\bz}$ with correlations given in \eqn{nu}. That is, 
$\bar{\alpha}^L_{n,\bx}={\alpha}^L_{n,\bx}$, while the equation for $\bar{\alpha}^R_{n,\bx}$
differs from that in \eqn{alar} just by letting $U^{ab}_{n-1,\bz} 
\to \bar U^{ab}_{n-1,\bz} $ in the r.h.s. Hence, the stochastic processes in the DA and the
CCA differ from each other only because the respective initial conditions,
$U^{\dagger}_0 = U^{\dagger}_A$ and $\bar{U}^{\dagger}_0 =\bar{U}^{\dagger}_A$, are different. Given this procedure, one iteratively constructs the evolved generating functional
\begin{align}\label{S12Lan}
 \big\langle {S}_{12}(\bx\bbx) \big\rangle_{\Delta Y} = \frac{1}{N_g}\,
 \big \langle \rmTr\big[\bar{U}^\pd_{N,\bbx} U^{\dagger}_{N,\bx}\big] 
 \big\rangle_{\nu}.
 \end{align}
By construction, this is a functional of the initial Wilson lines $U^{\dagger}_A$ and $\bar{U}^{\dagger}_A$.
Then one has to act with the production Hamiltonian at the initial rapidity $Y_A$ to emit the softest gluon, 
subsequently identify $\bar{U}_A=U_A$, and finally 
average over the target fields with the CGC weight function at $Y_A$. That is, one needs to evaluate
\beq
 \label{thereyougo}
 \int \mcal{D}U_A\,W_{Y_A}[U_A]\,
 H_{\rm prod}(\bk)[U_A,\bar{U}_A]
 \big\langle S_{12}(\bx\bbx) \big\rangle_{\Delta Y}
 \big|_{\bar{U}_A=U_A}.
 \eeq  
The action of $H_{\rm prod}$ on the generating functional 
involves  the linear combination of four terms, such as 
 \begin{align}
 \label{rrons}
 R^a_{A,\bu}\, \bar{R}^b_{A,\bw}\, 
 \big\langle S_{12}(\bx\bbx) 
 \big\rangle_{\Delta Y}
 \big|_{\bar{U}_A=U_A}
 = \frac{1}{N_g}\,
 \Big\langle 
 \rmTr \big[ 
 \big(R^b_{A,\bw} U_{N,\bbx}^\pd\big) 
 \big(R^a_{A,\bu} U^{\dagger}_{N,\bx}
 \big)\big]
 \Big\rangle_{\nu},
 \end{align}
where the subscript $A$ on the Lie derivatives emphasizes that they are acting on the functional dependence of the Wilson lines $U^{\dagger}_{N} $
and $\bar{U}_{N}$ upon their respective initial conditions $U_A^{\dagger}$ and $ \bar{U}_A$. 
Although conceptually well defined, this procedure seems to pose a serious difficulty in practice, since the use of functional initial conditions is not suited for numerics. However, this difficulty can be circumvented, as we now explain \cite{Iancu:2013uva}.

Namely, instead of following a {\em functional} Langevin process for the quantities $\bar{U}^{\dagger}_{n,\bx}$
and ${U}^{\dagger}_{n,\bx}$, it is more convenient to consider a {\em purely numerical} process, but for the
quantities ${U}^{\dagger}_{n,\bx}$ and $R^a_{A,\bu} U^{\dagger}_{n,\bx}$, where the second one
is {\em bi--local} in transverse coordinates. The results of such a process give us direct access
to the evaluation of the cross--section, via \eqn{rrons}.
 (Notice that $L^a_{A,\bu} U^{\dagger}_{n,\bx}$ does not require a separate calculation, in view of the relation $L^a_{A,\bu} = U^{\dagger ab}_{A,\bu} R^b_{A,\bu}$.) 
It is straightforward to obtain a recurrence formula by acting with $R^a_{A,\bu}$ on \eqn{un}. In fact, one
would rather use
 \beq
 \label{mcalr}
 \mcal{R}^a_{n,\bu\bx} = U^{\pd}_{n,\bx}
 R^a_{A,\bu} U^{\dagger}_{n,\bx},
 \eeq 
which is a member of the Lie algebra and it is a matter of a simple calculation to find
 \beq
 \label{rlang}
 \mcal{R}^a_{n,\bu\bx}
 =\rme^{\rmi \epsilon g \alpha^R_{n,\bx}}
 \mcal{R}^a_{n-1,\bu\bx}
 \rme^{-\rmi \epsilon g \alpha^R_{n,\bx}}
 -\frac{\rmi \epsilon g}{\sqrt{4\pi^3}}\,
 \rme^{\rmi \epsilon g \alpha^R_{n,\bx}}
 \int \dif^2 \bz
 \frac{\bx^i-\bz^i}{(\bx-\bz)^2}\,
 U^{bc}_{n-1,\bz} \nu^{ic}_{n,\bz}
 \big[T^b, \mcal{R}^a_{n-1,\bu\bz} \big],
 \eeq   
where the second term arises from the action of $R^a_{A,\bu}$ on the right phase in \eqn{un}. The initial condition for \eqn{mcalr} is not functional any more, but merely 
given by $\mcal{R}^a_{n,\bu\bx} = \rmi g \delta_{\bu\bx} T^a$. This initial condition is local in the transverse plane, however such a locality is immediately lost after the first evolution step, as evident in \eqn{rlang}.

To summarize, in order to calculate two-gluon production, we should first construct the Wilson lines $U_{\rm in}$ at the rapidity $Y_{\rm in}$ of the target nucleus, typically according to the MV model. Then from $Y_{\rm in}$ to the rapidity $Y_A$ of the `slowest' produced gluon we evolve according to the standard Langevin formulation of JIMWLK, 
as reviewed in Sect.~\ref{sec:lang}. Finally, from $Y_A$ until the rapidity $Y$ of the `fastest' produced gluon (which is similar to the projectile one in our setup), we also evolve the bi-local quantity $\mcal{R}^a_{n,\bu\bx}$, which measures the evolution of the color charge of an initial gluon via the successive
 emission of softer gluons. This procedure can be extended to an arbitrary number of partons in
 the final state, although the numerical manipulations become more and more involved  \cite{Iancu:2013uva}.


\providecommand{\href}[2]{#2}\begingroup\raggedright\endgroup

\end{document}